{

\documentclass{emulateapj} 

\slugcomment{\\Submitted to ApJ}

\usepackage{amsmath}
\usepackage{graphicx}
\usepackage{float}
\usepackage{amssymb,amsmath}
\usepackage{epsfig,floatflt}
\usepackage{subfigure}
\usepackage{color}

\def \ni{\noindent}
\def \th{\thinspace}
\def \dm1{{d$^{-1}$}}
\def \at{{\rm\char'100}}
\long\def\jumpover#1{{}}

\newcommand \etal{{\it et al.\ }}

\newcommand \viz{{\it viz.\ }}

\newcommand \half{{\frac 12 }}

\newcommand \approxgt{\,\raise2pt \hbox{$>$}\kern-8pt\lower2.pt\hbox{$\sim$}\,}
\newcommand \approxlt{\,\raise2pt \hbox{$<$}\kern-8pt\lower2.pt\hbox{$\sim$}\,}

\newcommand\aanda{A\&A}

\begin{document}

\title{Modulations in Multi-Periodic Blue Variables in the LMC}
\author{J. Robert Buchler\altaffilmark{1},
Peter R. Wood\altaffilmark{2} and
R. E. Wilson\altaffilmark{3}
}
\altaffiltext{1}{\ni Physics Department, University of Florida,
Gainesville, FL 32611, USA; 
buchler\at phys.ufl.edu}
\altaffiltext{2}{\ni Research School
of Astronomy \& Astrophysics, Australian National University, Canberra,
Australia,
wood\at mso.anu.edu.au}
\altaffiltext{3}{\ni Astronomy Department, University of Florida,
Gainesville, FL 32611, USA; 
wilson\at astro.ufl.edu}

\begin{abstract}

As shown by \cite{menni}, a subset of the blue variable stars in the Large
Magellanic Cloud exhibit brightness variability of small amplitude in the
period range 2.4 to 16 days as well as larger amplitude variability with
periods of $\sim $140 to 600 days, with a remarkably tight relation between the
long and the short periods.  Our re-examination of these objects has led to
the discovery of additional variability.  The Fourier spectra of 11 of their 30
objects have 3 or 4 peaks above the noise level and a linear relation of the
form $f_a = 2(f_b - f_L)$ among three of the frequencies.  An explanation of
this relation requires an interplay between the binary motion and that of a
third object.  The two frequency relations together with the Fourier amplitude
ratios pose a challenging modeling problem.

\end{abstract}



\section{Introduction}

A search for Be stars in the OGLE and MACHO databases by \cite{menni} uncovered
30 stellar objects that exhibit large amplitude low frequency light curve
modulations with frequencies in the range $f_L$ = 0.001 to  0.007 \th \dm1 (period
140 to 1000 d), in addition to one or two peaks in the $f_b$ = 0.05 to 0.50 \dm1
range (period 2 to 20 d).  Furthermore they found a
tight relation between the two peaks for all the objects, with the form 
\begin{equation}
f_b \approx 35 f_L.
\label{eq0}
\end{equation}
Analysis led to their suggestion that these objects are binary
stars seen under high orbital inclinations, in which a secondary star
overflows its Roche lobe onto a hotter primary star with the short term
periodicity being directly related to binary nature.
They then conjectured that the low frequency is due to an elliptical disk that
orbits the primary star. They leave open the
alternative of precession of a tilted disk around the primary.

 Interest in these objects extends beyond their specific peculiarities.
One can ask the general question of why there are very tight binaries with
dimensions similar to those of Main Sequence stars and periods of only a
few days, given that protostars are too large to fit within such systems.
Clearly some kind of orbit shrinkage is required. If the doubly periodic
binaries owe their unusual behavior to hierarchically distant third stars,
as suggested by newly detected periodicities of this paper, they are
likely to be interesting test objects for orbital evolution in multiple
systems. The role of third stars in close binary orbital evolution has
been examined over the past several decades - recently with a rapidly
developing literature (e.g. Mazeh \& Shaham, 1979, Kiseleva, Eggleton, \&
Mikkola, 1998, Eggleton \& Kiseleva-Eggleton, 2001, Fabrycky \& Tremaine,
2007). Simulations show that the orbit shrinkage mechanism of
Kozai cycles (Kozai, 1962) with tidal friction (KCTF), leads to production
of tight binaries and may account for virtually all ordinary binaries with
periods under 10 days. A third star can be effective in this regard at
remarkably large distance and low mass. Several of the cited papers have
excellent overviews of KCTF. This general dynamical issue arises also in
the problem of orbital migration in exoplanet systems (e.g. Mathis \& Le
Poncin-Lafitte, 2009). Unfortunately, observational probes that
potentially can establish relevant statistics are beset with severe
difficulties, as the distant companions may be too dim for detection or be
spatially unresolved, while reflex radial velocities of the inner binaries
are small, with inconveniently long periods. Although major progress has
recently been made on the observational side (Tokovinin, et al, 2006,
Pribulla \& Rucinski, 2006, D'Angelo, et al., 2006), detection of
companions and measurement of their orbits and masses remains very
difficult and degraded by selection effects. Therefore any new source of
information on possible companions to very close binaries needs to be
exploited for insights into the general problem, including formation,
orbital evolution, and stellar evolution.

When examining the MACHO database for low amplitude Cepheid variables
(\cite{buchler09})
we came across MACHO 77.7911.26, alias OGLE LMC SC3 274426,
in which we uncovered a relationship among the frequencies of the highest peaks.
This object is one of those analyzed by \cite{menni}.
We therefore had a closer look at all
their blue variables.
To our surprise we noticed that the frequency relation
\begin{equation} 
f_a = 2(f_b - f_L), 
\label{eq1} 
\end{equation} 
\noindent not mentioned by \cite{menni}, clearly exists among 11 of their 30 objects.

\begin{table*} 
\caption{The multi-periodic blue stars} 
\quad \quad \quad Frequencies $f$ in d$^{-1}$, amplitudes $A$ in mags.
\begin{center}{\scriptsize
\begin{tabular}{ r | r r r | r r r r | r r r r } 
\hline
\hline 
\noalign{\smallskip} 
Object & 
 $f_L$\quad & $f_a$\quad & $f_b$\quad & 
 $M_B$: $A_L$ & $A_a/A_L$ & $A_b/A_L$ & $A_s/A_L$ & 
 $M_R$: $A_L$ &$A_a/A_L$ & $A_b/A_L$ & $A_s/A_L$ \\
    \noalign{\smallskip} 
    \hline   
     \noalign{\smallskip}
77.7911.26  & 0.00166 & 0.12611 & 0.06469 & 0.0618 & 0.6363 & 0.2342 & 0.0798 & 0.0690 & 0.5514 & 0.1679 & 0.0577 \\
80.6469.95  & 0.00552 & 0.34850 & 0.17984 & 0.0438 & 0.2655 & 0.1746 & 0.2390 & 0.0461 & 0.2589 & 0.1798 & 0.1316 \\
1.4174.42   & 0.00439 & 0.25983 & 0.13427 & 0.1511 & 0.4197 & 0.0633 & 0.0066 & 0.1745 & 0.3385 & 0.0602 & 0.0151 \\
11.9592.22  & 0.00458 & 0.27956 & 0.14438 & 0.0406 & 0.2154 & 0.1927 & 0.1195 & 0.0592 & 0.1408 & 0.1086 & 0.1299 \\
80.6468.83  & 0.00702 & 0.41491 & 0.21451 & 0.0435 & 0.9404 & 0.1435 & 0.0522 & 0.0506 & 0.7367 & 0.1173 & 0.1119 \\
77.8033.140 & 0.00384 & 0.27843 & 0.14311 & 0.0933 & 0.2715 & 0.0981 & 0.1933 & 0.0991 & 0.2684 & 0.0894 & 0.1605 \\
212.15675.158& 0.00589& 0.38629 & 0.19903 & 0.1044 & 0.4655 & 0.1476 & 0.3281 & 0.1214 & 0.4682 & 0.1018 & 0.2657 \\
1.3686.53   & 0.00389 & 0.24923 & 0.12849 & 0.1426 & 0.2797 & 0.0282 & 0.1804 & 0.1680 & 0.2179 & 0.0543 & 0.1154 \\
11.9477.138 & 0.00486 & 0.30153 & 0.15570 & 0.0487 & 0.9897 & 0.1791 & 0.6932 & 0.0651 & 0.7421 & 0.0843 & 0.4615 \\
11.9475.96  & 0.00371 & 0.28704 & 0.14753 & 0.0612 & 0.5623 & --     & 0.3980 & 0.0796 & 0.4755 & 0.0257 & 0.2953 \\
79.5506.139 & 0.00548 & 0.37722 & 0.19164 & 0.0986 & 0.2481 & --     & 0.0049 & 0.1326 & 0.2907 & 0.0583 & 0.0112 \\
\noalign{\smallskip}
\hline   
\hline
\end{tabular}
}
\end{center}
\vskip 10pt
\label{tab1}
\end{table*}

\begin{figure*}
\begin{center}
 \includegraphics[width=2.in]{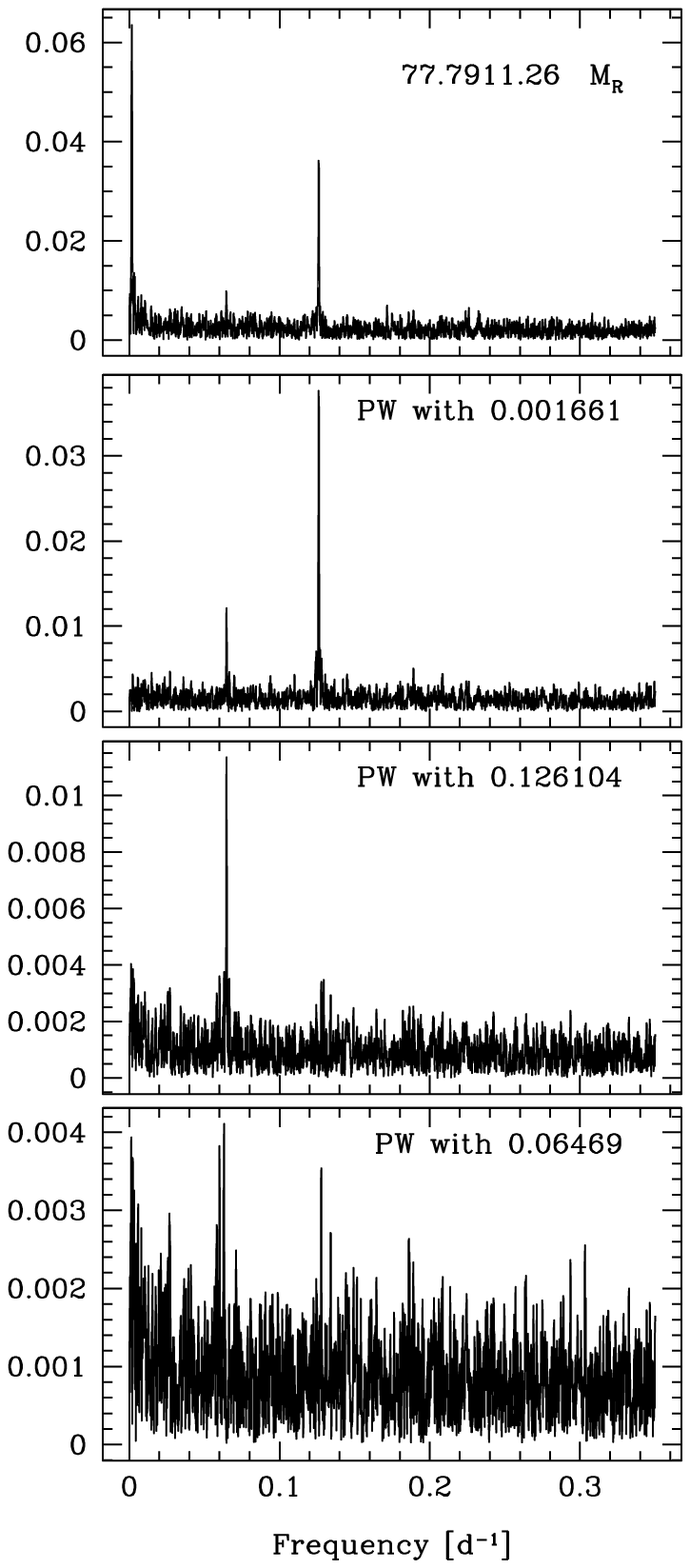}
 \includegraphics[width=2.in]{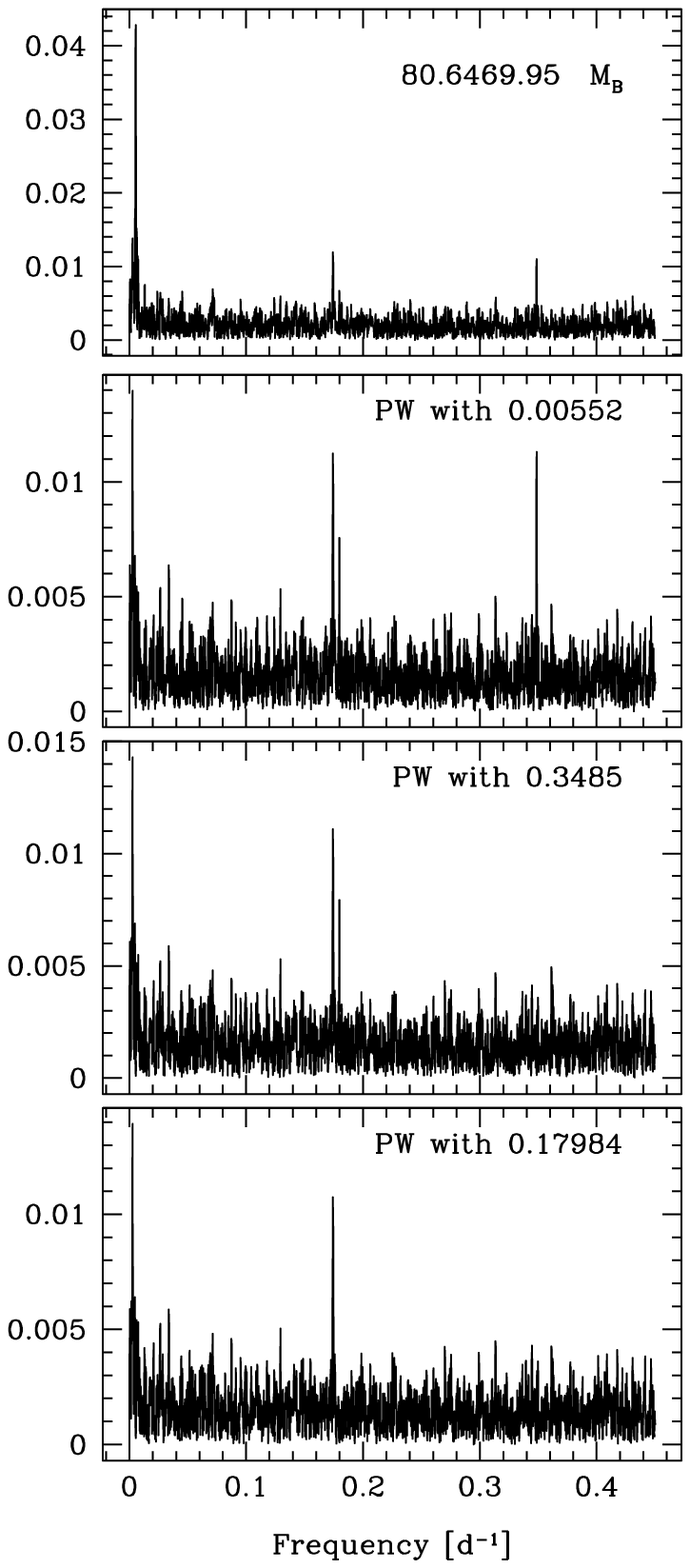}
 \includegraphics[width=2.in]{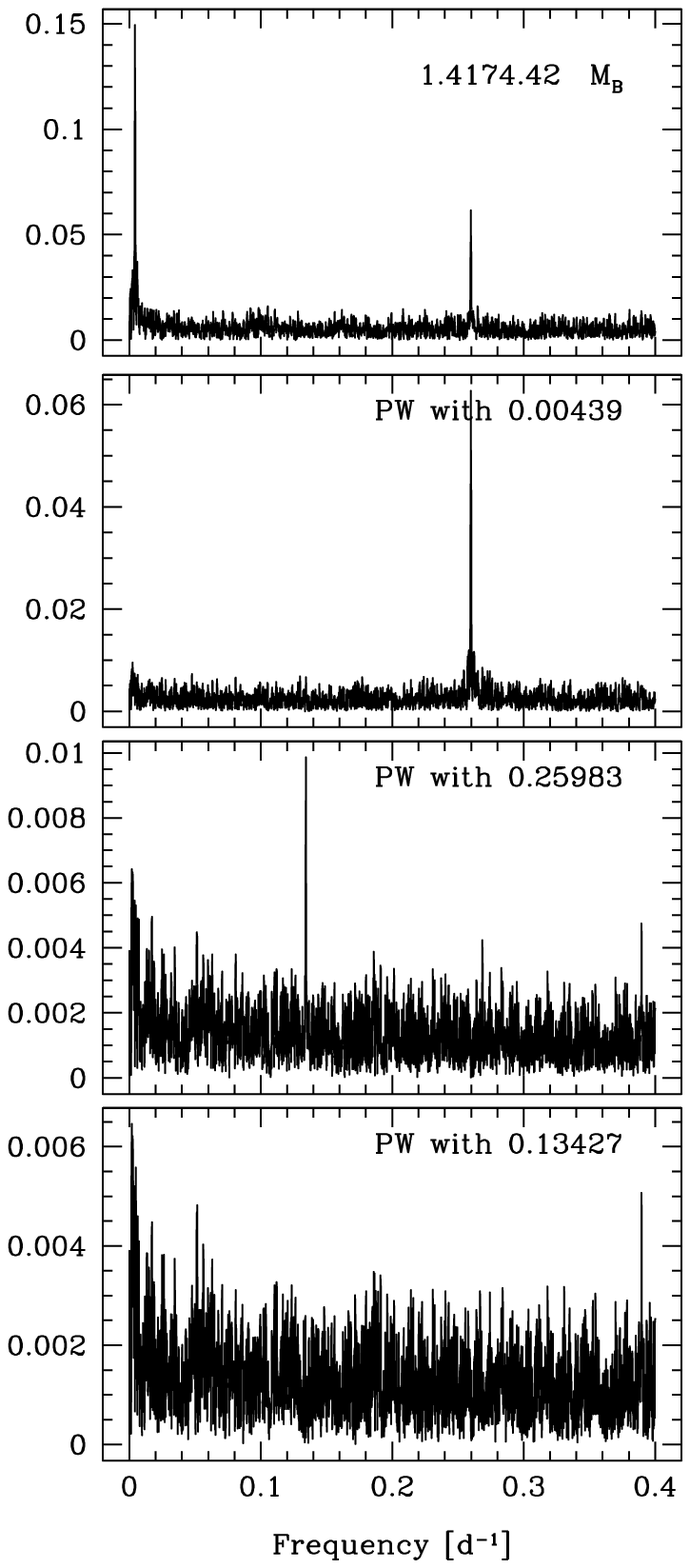}
\end{center}
\figcaption{
From top down: Amplitude Fourier spectra, followed by the spectra  
prewhitened successively with the 3 frequencies
$f_L$, $f_a$ and $f_b$.  
Note that $f_a = 2(f_b-f_L)$.
Left: Object 77.7911.26 in M$_R$ - 
the largest remaining peak at the bottom is the subharmonic $f_s = \half f_a$
at 0.06305 d$^{-1}$.
Middle: Object 80.6469.95 in M$_B$ -
the largest remaining
peak at the bottom is again the subharmonic $f_s$ at 0.17425 d$^{-1}$.
Right: Object 1.4174.42 in M$_B$; here the subharmonic  
is not visible above the noise level.  
} 
\label{figfs1}
\end{figure*}

Relation 2 between a pair of high frequencies and a much lower
frequency naturally suggests two approximately equal periodicities and
associated angular motions, leading to a model with a third star, as
discussed further in Section 3. However the third star characteristics
must be consistent with observed magnitudes and colors, so quantification
of the idea is not simple and is still in progress.

\begin{figure*}
\begin{center}
 \includegraphics[width=2.in]{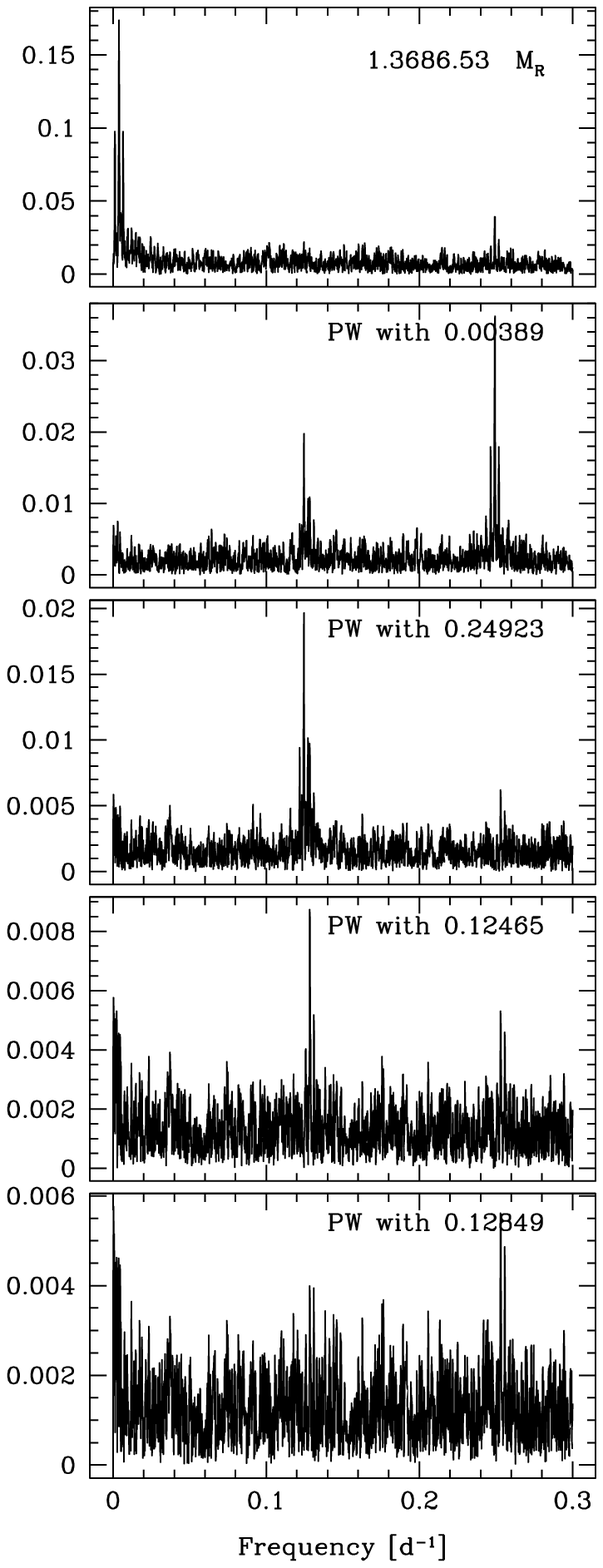}
 \includegraphics[width=2.in]{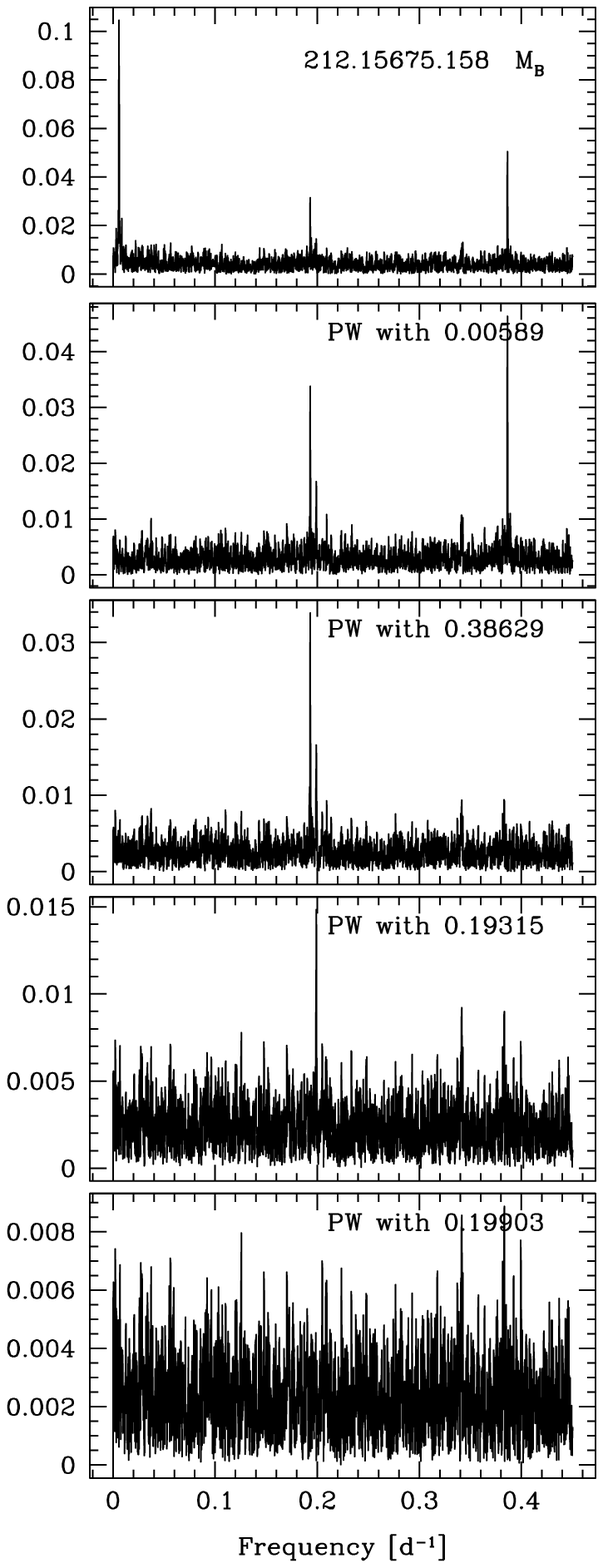}
 \includegraphics[width=2.in]{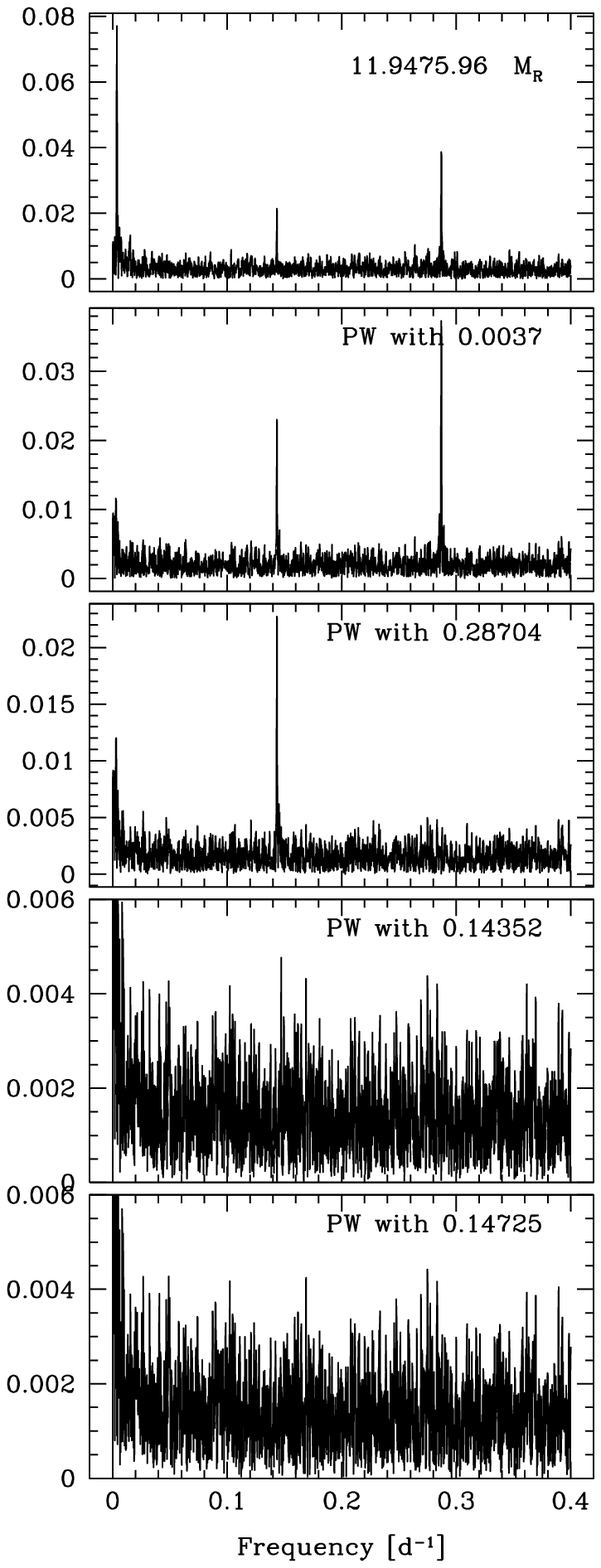} 
\end{center}
\figcaption{
From top down: Amplitude Fourier spectra  
prewhitened successively with the 4 frequencies
$f_L$, $f_a$, $f_s = \half f_a$ and $f_b$.  
Left:    1.3686.53 in M$_R$.
Middle:  212.15675.158 in M$_B$.  
Right: 11.9475.96 in M$_R$.  
}
\label{figfs2}
\end{figure*}

\section{Data Analysis}

Using the MUFRAN software \citep{mufran}  
we first performed a Fourier Transform 
over the frequency range 0 to 0.99 d$^{-1}$, after 
extirpolation\footnote{See \citet{pressryb} for an explanation of extirpolation.}
of the unequally spaced data.  The light curves were then
prewhitened with the dominant frequency and typically 3 harmonics, after the
dominant frequency had been estimated with a nonlinear least squares fit
\footnote{Prewhitening means subtracting a Fourier fit for a specified
frequency with some number of harmonics.}.  This procedure was repeated until
no significant frequencies remained.  Some of the data were also independently
checked with the Phase Dispersion Minimization method \citep{stell} via the
routine PDM in IRAF to confirm the detected MUFRAN frequencies.

Table~\ref{tab1} contains our Fourier results for the MACHO light curve bands
M$_B$ (blue) and M$_R$ (red).  Columns 2 to 4 show the frequencies of the
dominant peaks, $f_L$, $f_a$ and $f_b$, that satisfy Eq.~\ref{eq1}.  The M$_B$
amplitude $A_L$ of the long period modulation is in column 5 followed by the
relative amplitudes $A_a/A_L$, $A_b/A_L$ and $A_s/A_L$, where subscript s
hereafter refers to the subharmonic frequency $f_s=\half f_a$.  The
corresponding M$_R$ amplitudes appear in columns 9 to 12.

For the last two objects $f_b$ stands out above the noise level only in $M_R$.
The amplitude $A_L$ is always the largest one in both bands.  For about half
the objects $A_a$ is greater than $A_s$, while the reverse is true for the
others.  In two objects these amplitudes are reversed between M$_R$ and M$_B$.

Fig.~\ref{figfs1} illustrates Fourier spectra for objects
77.7911.26 in M$_R$, 80.6469.95 in M$_B$ and 1.4174.42 in M$_B$.  The top
panel shows the Fourier spectrum while the lower panels display
the spectra after successive prewhitenings, with the highest remaining peak and
its three harmonics.  The amplitudes decrease toward the noise level after
the three prewhitenings. The largest remaining peak is at
$f_s$ for MACHO 80.6469.95 and 77.7911.26, whereas it is hidden in the
noise for MACHO 1.4174.42.

When the light curves are prewhitened with the lowest frequency (and three
harmonics), much noise can remain at very low frequencies for some objects, in
addition to the ubiquitous yearly alias at $\approx$ 0.0027\th d$^{-1}$.  This
low frequency structure may dominate the higher frequency peaks, but we ignore
this junk power when labeling the frequency peaks in terms of amplitude.  For
MACHO 80.6469.95, the yearly alias remains quite prominent at $\sim$ 0.0027
d$^{-1}$.  When the subharmonic $f_s$ peak dominates over $f_a$ we do the
prewhitenings successively with peaks $f_L$, $f_a$, $f_s$ and finally with
$f_b$. Of course the second and third prewhitenings could have been done
jointly with $f_s$ and its harmonics.  Fig.~\ref{figfs2} presents other
examples, \viz objects MACHO 1.3686.53 in M$_R$, MACHO 212.15675.158 in M$_B$
and MACHO 11.9475.96 in M$_R$.

In MACHO 1.3686.53, $f_s$ and $f_b$ clearly stand out after
prewhitening, although they are hardly visible in the
original Fourier spectrum.
The $f_b$ peak is quite weak in MACHO 212.15675.158
but pops up visibly after 2 prewhitenings. MACHO
11.9475.96 shows a weak but real peak at $f_b$ after 2 prewhitenings that
is somewhat hidden by the large yearly peak
and low frequency noise.  We therefore changed the scale on the
bottom 2 panels (with the yearly peak off scale).  The $f_a$ peak is the 4th
largest if one disregards the low frequency noise.

\begin{figure}
\begin{center}
\includegraphics[width=2.1in]{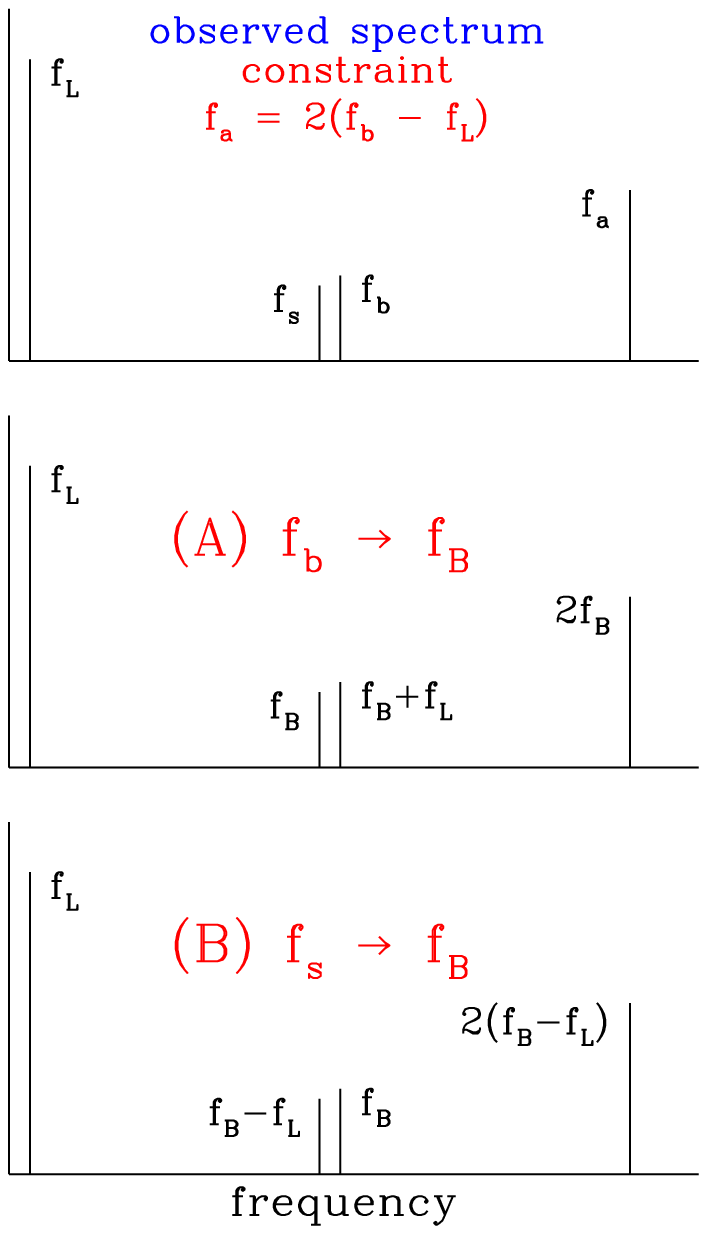}
\end{center}
\figcaption{Schematic Fourier amplitude spectrum.
Note that $f_s=\half f_a$.
Top diagram: Observed spectrum and frequency  constraint.
Middle and lower diagrams: Two possible sets of frequency assignments.
}
\label{fig:spectra}
\end{figure}


\section{Discussion}

We make the reasonable assumption, as in \cite{menni}, that one
component is a tight binary of frequency
$f_{B}$.  We also assume that frequency $f_L$ is associated with
a third object whose nature we leave unspecified at this stage.
Note that the light curves, prewhitened
with the frequency $f_L$ (which has the highest amplitude) and folded
with $f_a$, either suggest or are compatible with eclipses or
ellipsoidal variations.
Two sets of frequency assignments are possible for $f_b$,
Case A:
\begin{eqnarray}
f_s &\rightarrow & f_{B}\nonumber \\
f_a &=& 2 f_B\nonumber \\
f_b &=& f_B + f_L
\end{eqnarray}
or, Case B:
\begin{eqnarray}
f_b &\rightarrow & f_{B}\nonumber \\
f_s &=& f_B - f_L\nonumber \\
f_a &=& 2 (f_B-f_L).
\end{eqnarray}

These relations are displayed schematically in
Fig.~3.  We recall the observational conditions, namely
(Table~\ref{tab1}) that for all objects the Fourier amplitudes
satisfy $A_L > A_a$, that both $A_s$ and $A_b$ are smaller than $A_a$, and that
the ratio $A_s/A_b$ is close to unity, but can be above or below.

The combination frequency $f_B \pm f_L$ presumably is from radiation of the inner
binary that is reprocessed ("reflected") by the third object.
Note that in case A the motion associated with frequency $f_L$ is
retrograde, while in case B it is prograde.

In all of Table~1's 11 objects, $f_L$ has the highest amplitude and appears
to be reasonably sinusoidal (the harmonics have small amplitude). This means
that most of the light variation is generated by a third
object, separate from, but associated with the binary pair.  \citet{menni}
suggest that the third object could be an orbiting disk or a
precessing inclined disk.  Other possibilities involve tides in a third star on an
elliptical, long period orbit around the inner binary.

We prefer to separate the pure data analysis of these objects that has been
presented in this paper from a discussion of the physical nature and modeling
that we plan to address in a subsequent paper.

\section{Summary}

We have found a very clear relation of the form $f_b =
f_L + f_a/2$ among the dominant observed Fourier frequencies $f_L$, $f_b$ and
$f_a$ in 11 of the 30 LMC blue variables discussed in \cite{menni}.
This relation is found in both MACHO bands, except for 2 objects where the
$f_b$ peaks seem to be hidden in the noise in $M_B$.  
Accordingly, strong observational constraints apply to all 11 objects,
namely (a) the frequency ratio $f_b/f_L\sim 35$ (or $f_a/f_L\sim 70$) found by
\cite{menni}, (b) the linear 3-frequency relation of this paper, (c) the amplitude
ratios of the 4 observed peaks (Table~\ref{tab1}), and (d) the absence of
other significant harmonics or peaks in the Fourier spectra.  These conditions
pose an interesting and serious challenge for development of a physical model
for these 11 objects.

\begin{acknowledgements}

It is a pleasure to thank Zoltan Kollath for providing his MUFRAN software.
JRB has enjoyed the hospitality of Mount Stromlo Observatory where this work
was started.  This work has been supported by NSF grants
AST~0707972, AST~0307561 and OISE04-17772 at the University of Florida.

\end{acknowledgements}


\end{document}